\documentclass[letter]{aa} 

\usepackage{graphicx}
\usepackage{txfonts}

\begin{document}

   \title{ Unraveling the birthplaces of NGC\,2070's massive stars,  tracked with MUSE and revealed with JWST}

   \author{N. Castro
          \inst{1,2}
          }

                  \institute{Leibniz-Institut für Astrophysik Potsdam (AIP), An der Sternwarte 16, 14482, Potsdam, Germany
                  \email{ncastro@aip.de}
           \and
           Institut für Astrophysik, Georg-August-Universität, Friedrich-Hund-Platz 1, 37077 Göttingen, Germany
}

\abstract{

The formation of massive O-type stars cannot be simply explained as a scaled-up version of the accretion mechanisms observed in lower-mass stars. Understanding these processes necessitates systematic studies of their early stages, which are challenging to identify. Forming massive stars remain embedded in their dense nursery clouds, and IR instruments with high spatial resolution capabilities are needed to better observe them. Despite these challenges, MUSE optical observations of the massive cluster NGC 2070 successfully detected potential star-forming regions through spatially resolved electron density maps. To further explore these regions, the \textit{James Webb} Space Telescope (JWST) utilized its NIRCam and MIRI instruments to penetrate optically obscured areas. This study examines two specific regions in the southeast part of the NGC 2070 MUSE density map, where tracks of highly dense point sources were identified. NIRCam, partially overlapped with MIRI, resolved these MUSE findings, revealing a procession of stellar point sources in the projected images. The detections are associated with elongated clouds, suggesting greater proper motions compared to the surrounding interstellar medium. These findings may indicate the presence of runaway candidates in the early stages of their evolution that are following common escape routes. This would support the notion that dynamical ejection is an efficient mechanism for the formation of massive runaway stars during early stages and likely plays a significant role in the origin of O-type field stars. However, additional data are required to confirm this scenario and rule out other ionizing feedback mechanisms, such as those observed in the formation of pillar-like structures around HII regions in the Milky Way. MUSE electron density mapping effectively captures the complexity of NGC 2070's interstellar medium and highlights targets for subsequent spectroscopic follow-ups, as demonstrated by the JWST data in the two fields studied.

}

\maketitle %

\nolinenumbers

\section{Introduction}

Massive O-type stars play a central role in shaping the Universe as they act as primary sources of heavy elements and  UV radiation. O-type stars, characterized by their high mass (M/M$\odot > 16$), are relatively rare and consume their hydrogen fuel at a much faster rate compared to lower-mass stars. However, the formations and evolution of galaxies are intricately linked to these stars, both chemically and dynamically. Throughout their relatively short lifetimes, massive stars undergo intense stellar winds and expel substantial amounts of material through massive outflows \citep{2012ARA&A..50..107L}. Finally, their spectacular explosive deaths have a profound impact on the surrounding interstellar medium (ISM), influencing its composition and structure \citep{1998ARA&A..36..189K}.

The formation of massive O-type stars remains a puzzling phenomenon, and various scenarios have been proposed to explain their origins. These include monolithic collapse with an  accretion disk, competitive accretion, stellar collisions and/or mergers \citep{2002ApJ...569..846Y,2001MNRAS.323..785B}. It is widely acknowledged that the formation of massive stars is not simply an up-scaled version of the processes involved in the formation of lower-mass stars. To unravel the mysteries surrounding their formation, additional data are crucial \citep{2007ARA&A..45..481Z}. However, the detection and analysis of the initial stages of massive star formation pose significant challenges due to the combination of insufficient spatial resolution and the fact that these stars spend approximately 15\% of their lifetimes embedded within their highly dense nursery clouds \citep{2002ASPC..285...40H,2005ApJ...631L..73S}.

\begin{figure*}
        \centering
        \includegraphics[angle=0,width=\textwidth]{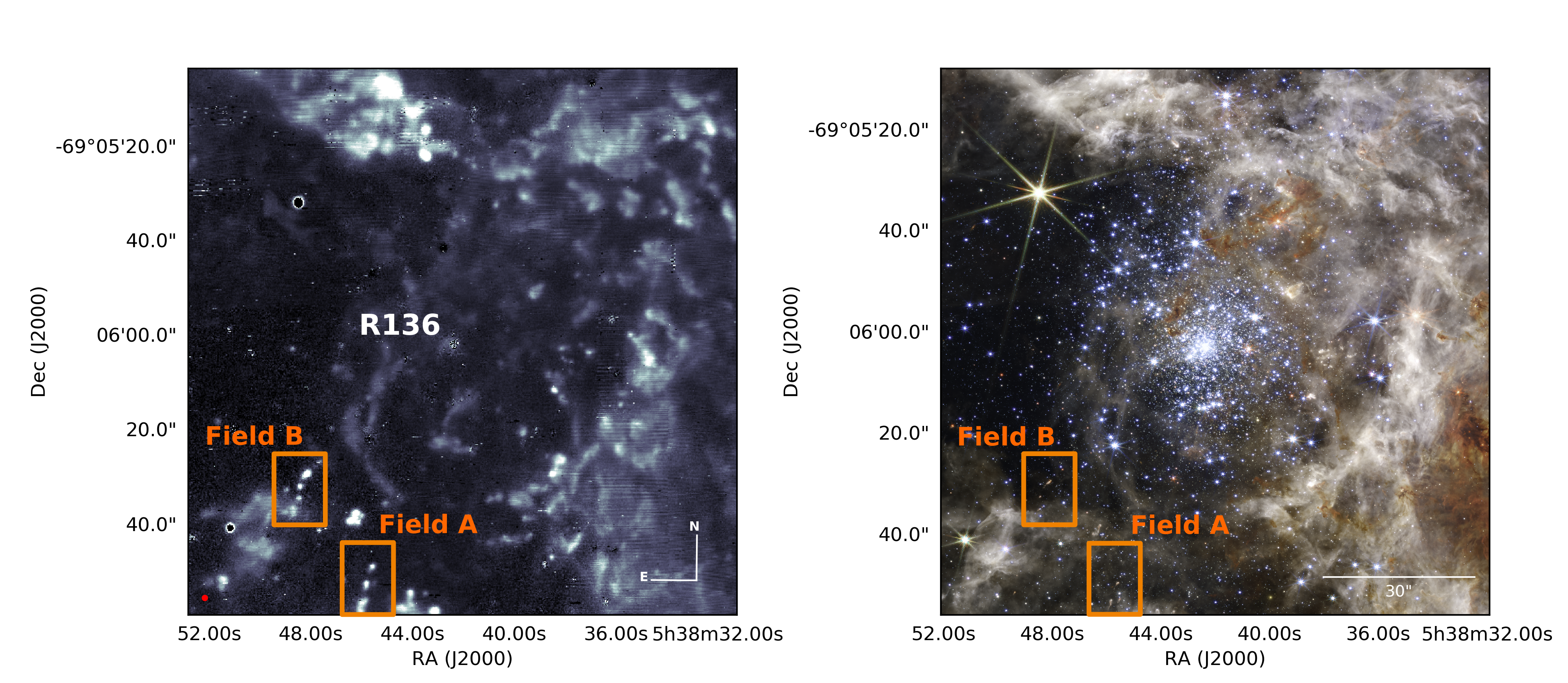}

        \caption{The NGC\,2070 massive stellar cluster. Left: Electron density map of NGC 2070 based on the [\ion{S}{ii}] $\lambda$6717/6731 ratio \citep{2018A&A...614A.147C}. The core of NGC\,2070, known as the R136 cluster, is labeled for reference. The image quality of MUSE, with an approximate resolution of 1", is indicated by the red dot in the bottom-left corner. Right: Color-composite image of the cluster obtained using the JWST NIRCam. The regions of interest explored in this study, fields A and B, are  marked on both panels.}

        \label{fig:all}
\end{figure*}

Infrared (IR) and submillimeter instruments have proven to be powerful tools for overcoming these challenges and investigating tightly embedded massive clusters \citep{2006IAUS..232..324Z,2007ApJ...665..478K}. However, optical integral field spectroscopy can also play a crucial role in studying the evolution of massive stars and identifying star-forming candidates. \cite{2018A&A...614A.147C}, using data from the Multi Unit Spectroscopic Explorer (MUSE) integral field spectrograph  \citep{2014Msngr.157...13B}  of the massive cluster NGC\,2070, provided the largest and most comprehensive spectroscopic analysis of the massive stellar population within the Tarantula Nebula \citep[see also][]{2021A&A...648A..65C}. That work not only revealed the kinematics, electron temperature, and extinction maps of the ISM surrounding the stellar sources but also used the electron density map, based on the [\ion{S}{ii}]$\,\lambda$6717/6731 ratio \citep{1984MNRAS.208..253M}, to identify areas in the cluster where new star formation is likely occurring. Some of these dense regions corresponded to previously reported young stellar candidates \citep{2009ApJS..184..172G}, such as the large region in the northeast of the MUSE field reported by \cite{2002AJ....124.1601W}. In the southeast, slightly separated from the areas dominated by ISM and H$\alpha$ emission in optical observations, the MUSE density map revealed several isolated and compact regions that exhibit apparent ordered patterns in the projected sky (see the left panel of Fig.~\ref{fig:all}). The electron density values in these regions were notably higher than those in nearby areas and the average measurement in NGC\,2070 \citep{2010ApJS..191..160P,2018A&A...614A.147C}. Table~\ref{tab:targets} presents the targets shown in the two panels of Fig.~\ref{fig:all} and the corresponding electron densities derived from the [\ion{S}{ii}]$\,\lambda$6717/6731  ratio \citep{2018A&A...614A.147C}. These organized sequences of star-forming candidates could in theory help us unravel the formation and potential kinematics of massive clusters during their early stages. However, the spectra extracted from the MUSE datacube did not reveal any stellar photospheric features of the embedded objects. Moreover, the catalog published by Castro et al. did not report any objects at the positions marked in fields A and B (Fig.~\ref{fig:all}) and listed in Table~\ref{tab:targets}. While the [\ion{S}{ii}]$\,\lambda$6717/6731  optical ratio can unambiguously identify pre-main-sequence massive star candidates, these sources are too faint and obscured to be studied using the current optical MUSE data.

In this study, I aim to resolve two of the particularly interest regions that were identified using MUSE, by utilizing the high spatial resolution IR  data obtained from the Near-Infrared Camera (NIRCam) and Mid-Infrared Instrument (MIRI) on the \textit{James Webb} Space Telescope (JWST; \citealt{2006SSRv..123..485G}). After discussing the data used in this work in Sect. \ref{data}, I demonstrate in Sect. \ref{track} the superior capabilities of NIRCam and MIRI for dissecting the origin of MUSE star-forming detections.  In Sect.~\ref{spec}, I explore the kinematics of the [\ion{S}{II}]$\,\lambda\lambda\,6716,6731$ emission lines, mapped with MUSE, after filtering the average ISM background. Finally, in Sect. \ref{conclusion}, I discuss the findings of this study and outline potential future directions for continuing this work.

        \begin{table}
			\caption{Coordinates and electron densities (n$_e$) of the sources depicted in Figs.~\ref{fig:area1} and \ref{fig:area2}. }

                \centering
                \begin{tabular}{cccccc}
                \hline\hline
                
                ID & RA & DEC & n$_e$ & vr. & err.\\
                 & hh:mm:ss & dd:mm:ss & cm$^{-3}$ & km$\,$s$^{-1}$ & km$\,$s$^{-1}$\\
        
                \hline
                A1 & 05:38:45.58 & -69:06:49.4 & 488 & -35.4& 6.9\\
                A2 & 05:38:45.80 & -69:06:52.9 & 590 & -24.1&5.3 \\
                A3 & 05:38:45.97 & -69:06:57.1 & 670 &-41.5 &2.6 \\
                A4 & 05:38:46.66 & -69:06:52.1 & 107 & 11.5&15.7 \\
                B1 & 05:38:47.65 & -69:06:27.3 & 182 &-27.2  &  10.3\\
                B2 & 05:38:48.09 & -69:06:29.5 & 1132& -10.3& 0.8 \\
                B3 & 05:38:48.41 & -69:06:32.3 & 517 & -12.3 &2.4 \\
                B4 & 05:38:48.44 & -69:06:34.8 & 392 & -133.7& 34.2\\
                B5 & 05:38:48.96 & -69:06:33.0 & 208 &-15.5 & 2.4\\
                \hline
                \end{tabular}

                \tablefoot{The electron densities are calculated as averages within a 1" aperture centered on each target, using the MUSE density map (Fig.~\ref{fig:all}). The average n$_e$ of the sky surrounding the targets, which is approximately 180 cm$^{-3}$, has been subtracted from the measurements. The average radial velocities (vr.) and dispersions (err.) of the residual [\ion{S}{ii}]$\,\lambda\lambda\,$6716, 6731 lines, following background filtering (as described in Sect.~\ref{spec}), within a 1" aperture, are listed in the last two columns.}

                \label{tab:targets}
        
        \end{table}

\section{Data}
\label{data}

As part of the MUSE Science Verification program, NGC\,2070 was observed by the Very Large Telescope in August 2014. The observation consisted of four overlapping fields, each covering an area of 1'$\times$1', and effectively mapped the core of the cluster. The data were processed using the MUSE pipeline, employing the esorex\footnote{https://www.eso.org/sci/software/cpl/esorex.html} recipes developed by \cite{2012SPIE.8451E..0BW}. To enhance the signal-to-noise ratio and mitigate the impact of cosmic rays and instrumental artifacts, the individual exposures were combined. A comprehensive overview of the observations is presented  in \cite{2018A&A...614A.147C,2021A&A...648A..65C}.

The JWST NIRCam and MIRI images of NGC\,2070 are part of the Webb Early Release Observations (ERO) conducted under proposal ID 2729  (PI K. Pontoppidan). For this study, I utilized the color-composite images provided by the ERO team. The NIRCam colored image was created by combining data from the F090W, F200W, F335M, and F444W NIRCam filters. The MIRI composite image was generated using the F770W, F1000W, F1280W, and F1800W filters. 

Both the MUSE and JWST images underwent astrometric calibration using \textit{Gaia} \citep{2023A&A...674A...1G} to ensure that they share the same coordinate system. This  calibration process guarantees a high level of accuracy, facilitating a reliable  comparison between the MUSE and JWST data. Moreover, it allows for the unambiguous identification  of the  MUSE sources in the JWST images. The field of view of the MUSE instrument toward NGC 2070 is depicted in Fig.~\ref{fig:all} and covers an area of 2'$\times$2'. Additionally, this figure displays the corresponding IR counterpart observed by NIRCam.

\section{Tracking MUSE star-forming regions with JWST}
\label{track}

\begin{figure*}[!h]
        \centering
        \includegraphics[angle=0,width=\textwidth]{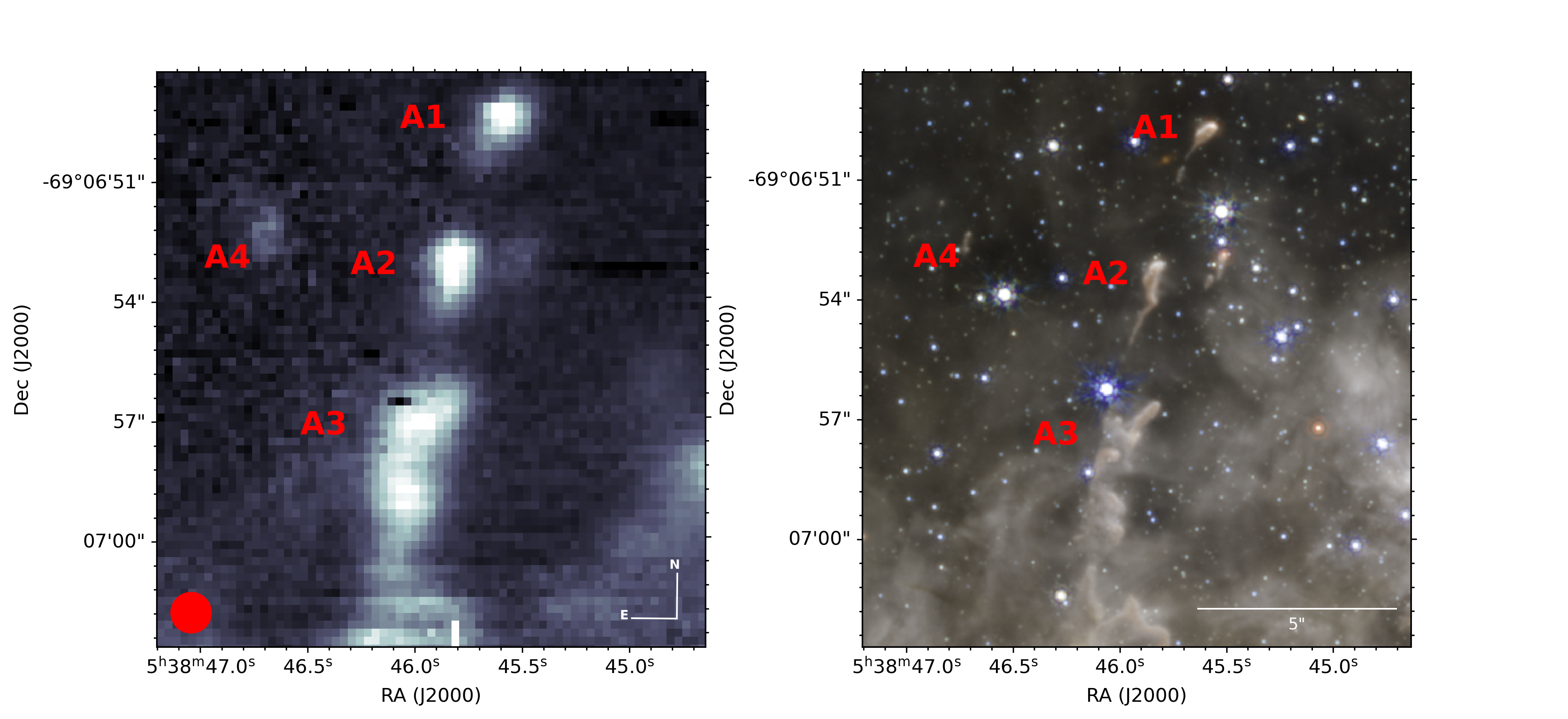}
        \caption{Zoom in on field A. Some of the most prominent targets in the field are labeled. For a comprehensive description of the two panels, refer to Fig.~\ref{fig:all}.  }
        \label{fig:area1}
\end{figure*}

\begin{figure*}
        \centering
        \includegraphics[angle=0,width=\textwidth]{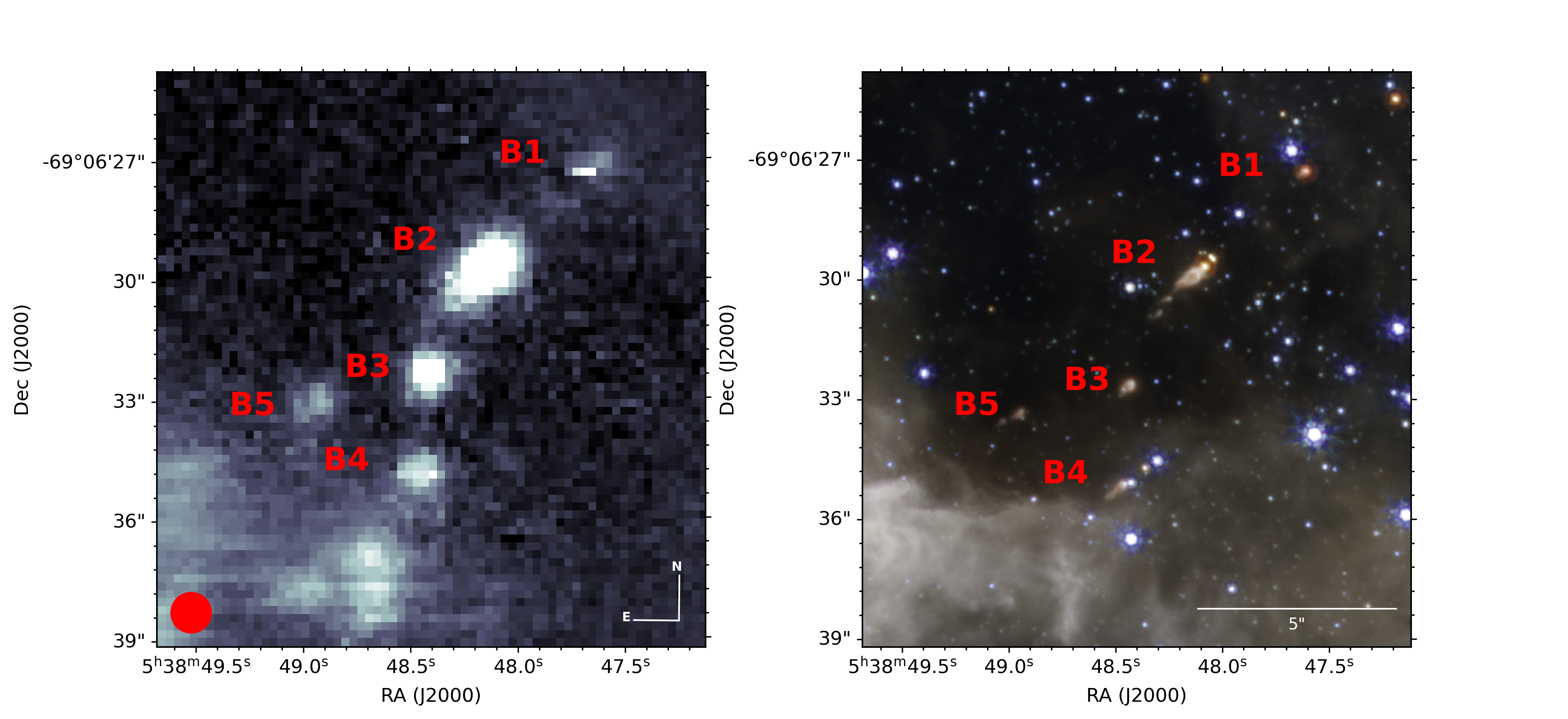}
        \caption{Zoom in on field B. Some of the most prominent targets in the field are labeled. For a comprehensive description of the two panels, refer to Fig.~\ref{fig:all}. }
        \label{fig:area2}
\end{figure*}

Mapping the regions highlighted with MUSE using optical spectroscopy is challenging. Exploring these regions necessitates the use of instruments with higher spatial resolution and IR capabilities. The JWST has conducted observations of 30Dor in the IR range using NIRCam. The remarkable color composition of NGC\,2070, constructed from NIRCam observations in four filters, showcases the exceptional spatial resolution and sensitivity offered by JWST\footnote{https://webbtelescope.org/contents/media/images/2022/\\041/01GA76MYFN0FMKNRHGCAGGYCVQ?news=true}. The NGC 2070 color composition (also visible in the right panel of Fig.~\ref{fig:all}) serves as a tool for investigating the star-forming candidates identified through MUSE and the [\ion{S}{ii}] $\lambda$6717/6731 ratio. Figures~\ref{fig:area1} and \ref{fig:area2} provide a closer look at  two intriguing regions within NGC\,2070. The MUSE electron density map is displayed  in the left panels, and the NIRCam equivalent to the right. As anticipated, JWST significantly enhances the spatial resolution of the MUSE data (by approximately 1"). While MUSE reveals densely concentrated regions, JWST portrays point sources enveloped by elongated nebulosities, indicating potential larger proper motions compared to the average local ISM kinematics of NGC\,2070. Evidence of these apparent projected movements can be seen in Figs.~\ref{fig:area1} and \ref{fig:area2}, which show our targets following a preferred path across the sky toward the projected center of the cluster. It should be noted that this does not necessarily imply that they are physically moving toward the central R136 cluster.

\begin{figure*}
                \centering
                \includegraphics[angle=0,width=\textwidth]{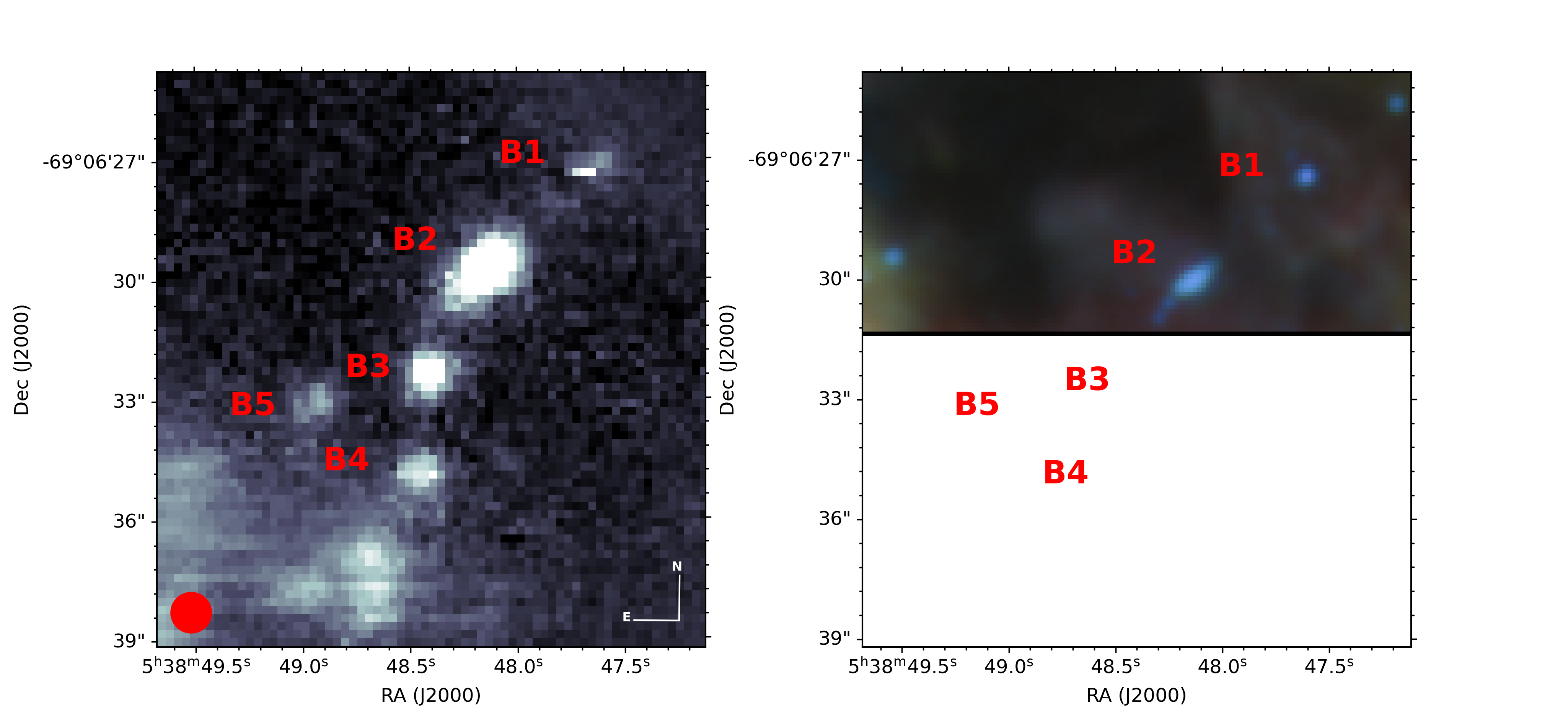}
                \caption{Zoom in on field B. I present an updated view where the color-composite JWST NIRCam image from Fig.~\ref{fig:area2} has been replaced with the color-composite JWST MIRI image.  Note that the current JWST MIRI image of NGC\,2070 only partially covers field B.    }
                \label{fig:area2_mir}
\end{figure*}

Further data at mid-IR wavelengths from JWST MIRI are available. Regrettably, the  color composition released as part of the Webb ERO\footnote{https://webbtelescope.org/contents/media/images/2022/\\041/01GA77BCCQDQ8JZ0D5FCN70QHK?news=true} does not encompass the targets depicted in Fig.~\ref{fig:area1}, and only partially covers those in Fig.~\ref{fig:area2}. Figure~\ref{fig:area2_mir} presents a comparison between the MUSE electron density and MIRI data.  The object identified as B1 has been observed, and its mid-IR counterpart is aligned with the elongated tail, as revealed by NIRCam and, at lower spatial resolution, MUSE electron density map.

We must consider the possibility that a fortuitous combination of obscured clouds and gaps in the line of sight may be deceiving us \citep[e.g.,][]{1947ApJ...105..255B}, creating the impression of point sources moving in a specific direction.  \cite{2018A&A...614A.147C} also conducted a mapping of the extinction in NGC\,2070 based on the H$\beta$/H$\alpha$ ratio. The extinction map reveals areas, not necessarily dominated by ISM emission, that exhibit higher levels of extinction in the southwest region (Fig. 8 in \citealt{2018A&A...614A.147C}). However, this is not observed in the vicinity of the objects discussed here in the southeast part of the MUSE map. Additionally, neither the near-IR image nor the partial mid-IR observation supports the presence of optically obscured clouds along the line of sight in fields A or B.

\section{[\ion{S}{ii}]$\,\lambda\lambda\,6716,6731$ MUSE radial velocity maps }
\label{spec}

Due to the MUSE spectrograph's relatively low spectral resolution of approximately 60\,km\,s$^{-1}$, analysis of the radial velocities of the candidates in the two fields  is limited. Nevertheless, the broader and irregular shapes observed in the  [\ion{S}{ii}]$\,\lambda\lambda\,6716,6731$ lines, particularly in specific areas of the fields, suggest the presence of blended components that can provide insights into the kinematics.

\begin{figure*}
        \centering
        \includegraphics[angle=0,width=\textwidth]{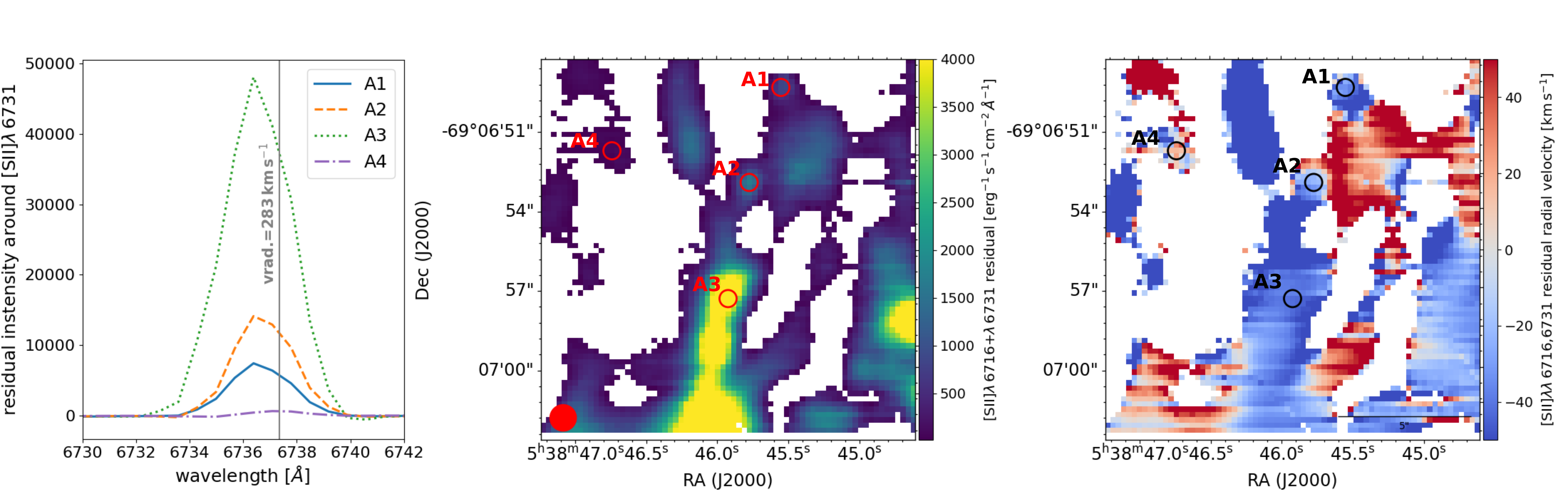}

        \caption{ Residual map in field A after \textsc{nebulisier} background filtering. Left panel: Residual [\ion{S}{ii}]$\,\lambda\,6731$ profiles extracted for each labeled target in field A with a 1" aperture. The gray line marks the average radial velocity of the ISM in the field. Middle panel: Residual intensity map of [\ion{S}{ii}]$\,\lambda\lambda\,6716,6731$. Right panel: Average radial velocities of [\ion{S}{ii}]$\,\lambda\lambda\,6716,6731$ lines. The targets listed in Table~\ref{tab:targets} are identified in both maps.}
        \label{fig:rv1}

        \includegraphics[angle=0,width=\textwidth]{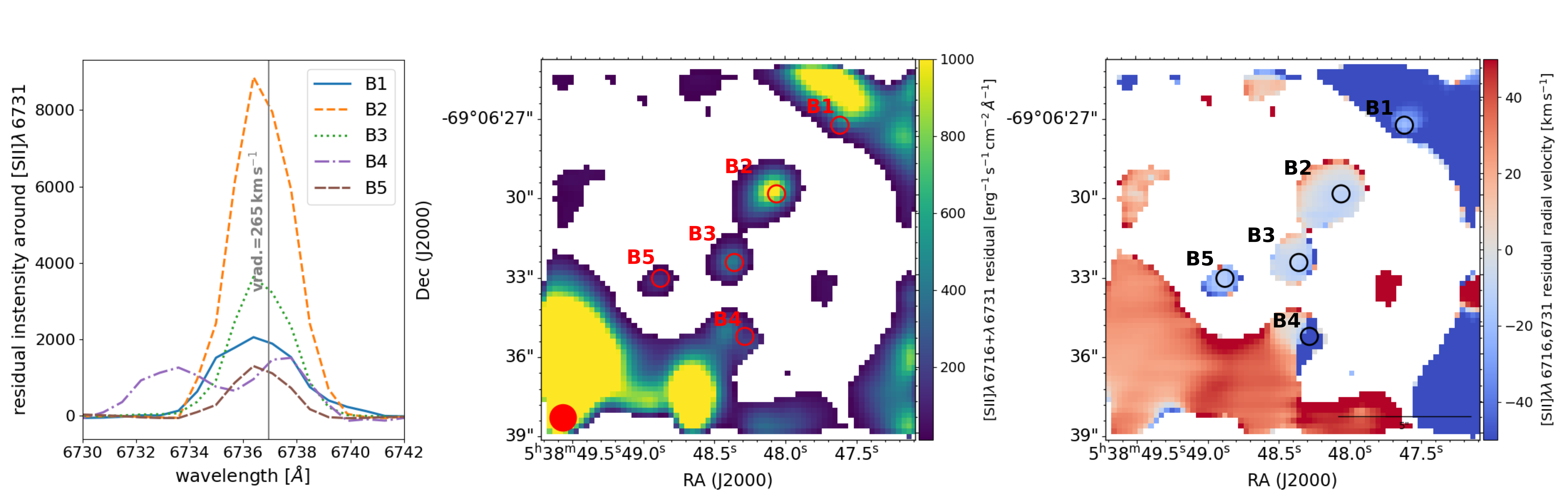}

        \caption{ Residual map in field B after \textsc{nebulisier} background filtering. For a comprehensive description of the panels, refer to Fig.~\ref{fig:rv1}. }
        \label{fig:rv2}
\end{figure*}

The ISM surrounding fields A and B was modeled using the \textsc{nebulisier} code\footnote{http://casu.ast.cam.ac.uk/surveys-projects/software-release/background-filtering} \citep{irwin2010nebulosity} and subsequently subtracted. The intensity residual maps of [\ion{S}{ii}]$\,\lambda\lambda\,6716,6731$ are presented in Figs.~\ref{fig:rv1} and \ref{fig:rv2}, highlighting regions where multiple contributions to the [\ion{S}{ii}] profiles were identified. The targets examined in this study are indicated within these residual maps in Figs.~\ref{fig:rv1} and \ref{fig:rv2}. In field A (Fig.~\ref{fig:rv1}), the targets are discernible in the intensity maps, but the field exhibits a complex distribution. The IR counterpart of NIRCam shown in Fig.~\ref{fig:area1} suggests the ISM in field A is intricate. Conversely, the view of field B shown in Fig.~\ref{fig:rv2} is clearer, with B2, B3, and B5 unmistakably separated from the background model using \textsc{nebulisier}. B1 and B4 are also detected but are mixed with other prominent residual components.

Figures~\ref{fig:rv1} and \ref{fig:rv2} depict the radial velocities of [\ion{S}{ii}]$,\lambda\lambda,6716,6731$ lines following background filtering. While the 2D radial velocity maps of field A do not provide individual source separation, as indicated by the density map and JWST observations, they reveal a noteworthy pattern: the primary filament enclosing the targets consistently exhibits a blueward shift in comparison to the average radial velocity of the field. The left panel of Fig.~\ref{fig:rv1} illustrates the average [\ion{S}{ii}]$\,\lambda\,6731$ profiles extracted from a 1" aperture around each target, showing their systematic displacement toward the blue end relative to the field's average value. Field B offers a clearer view of the targets resolved by JWST. All five objects (B1 to B5) exhibit systematic blueward velocities compared to the region's average velocity. The average radial velocities and dispersions extracted from the two fields are summarized in Table~\ref{tab:targets}.

The residual maps of both fields reveal a complex image around the targets unveiled by MUSE and JWST. Notably, in field B, emissions exhibit strong redward and blueward shifts relative to the field's average velocity. For instance, the average profile of B4, as seen in the left panel of Fig.~\ref{fig:rv2}, displays a double profile with an apparent component moving at approximately -150 km\,s$^{-1}$. While further exploration of these regions and an expansion of the studied areas could provide us with a clearer understanding of the stellar activity in fields A and B, such endeavors fall beyond the scope of this work and should be explored in subsequent studies.

\section{Discussion and future steps}
\label{conclusion}

Early pioneering works \cite[e.g.,][]{2004ApJ...603..531R} have shown that integral field spectrographs are a mature technology. MUSE is the testimony of this. The large field of view and spatial resolution has allowed us to explore dense clusters such as NGC\,2070 and systematically resolved highly star-forming pockets, despite them being immersed in their circumstellar material. The MUSE electron density map has guided us toward these star-forming regions and prevented them from going unnoticed, making them valuable targets for further analyses.

This work focuses on two regions where the MUSE density map detects aligned sequences of quasi-punctual sources in the project image. NIRCam and MIRI reveal processions of stellar point sources in both fields surrounded by apparent elongated clouds, suggesting differential kinematics with respect to the average ISM in the fields.

Dissection of the radial velocities of these components from the average ISM is hampered by the low spectral resolution of MUSE. However, [\ion{S}{ii}]$\,\lambda\lambda\,6716,6731$ profiles exhibit additional broadening beyond the expected instrument resolution. After applying background filtering using the \textsc{nebulisier} code, the residual intensity maps of [\ion{S}{ii}]$\,\lambda\lambda\,6716,6731$ highlight the targets in fields A and B, which consistently show blueward shifts relative to the average radial velocity in each field. The spectroscopic analysis of the background-subtracted MUSE data supports the runaway scenario; however, a higher spectral resolution is necessary to map the distinct overlapping components.

The observed patterns exhibited by the objects in fields A and B suggest potential escape routes for the forming stars, leading them away from their birthplaces. This discovery introduces new prospects for identifying runaway stars in their very early stages and provides valuable insights into the origins of isolated massive stars \citep{2020ApJ...903...42V}. Considering the young ages of these stars, it seems more plausible to attribute their displacement from the parent cluster to dynamical ejection rather than a supernova kick \citep{1961BAN....15..265B}. The \textit{Chandra} X-ray maps, as presented in the work by \cite{2006AJ....131.2164T}, reveal no discernible counterparts in the vicinity of these sources, providing no evidence of previous supernova events or the binding of any of these sources to a compact object post-supernova.

Previous studies  \citep{2008A&A...490.1071G,2011A&A...535A..29G} have suggested that dynamical ejection resulting from intense interactions with other stars is the primary mechanism responsible for massive stars leaving their birthplaces. Furthermore, a significant portion, if not all, of the isolated massive stars detected in these fields  \citep{2012MNRAS.424.3037G} have been identified as runaway stars, likely generated through dynamic encounters during the early stages of their formation \citep{2016A&A...590A.107O,2018ApJ...867L...8O,2020ApJ...903...43D}. Fields A and B may provide a privileged opportunity to witness the initial phases of dynamical ejection involving several stars that are still in the process of forming.

Field A's birthplace appears to be primarily situated to the south, extending beyond the region covered by the MUSE observation. In contrast, field B presents a more intriguing scenario. Following the trajectory indicated by the stars in Fig.~\ref{fig:all} leads us toward a red supergiant and the prominent ISM clouds enveloping it. \cite{2018A&A...614A.147C} also reported higher electron temperatures in these regions, although it is worth noting that it was not possible to determine the electron temperature at the target positions. The tracing and exploration of the birthplaces will be addressed in forthcoming studies.

Nonetheless, without additional data, I cannot rule out that we are witnessing the result of feedback mechanisms similar to those observed around massive stars in the Milky Way \citep[e.g.,][]{2002A&A...393..251S,2004ApJS..154..385R}. The resemblance of field A to known pillar-like structures in the Milky Way \citep[e.g.,][]{2015MNRAS.450.1057M} points to a similar scenario in which turbulence in molecular clouds, shaped by radiation emitted from surrounding massive stars, may produce these structures \citep{2010ApJ...723..971G}. The apparently isolated sources found in field B suggest the possibility of runaway objects. However, the current MUSE spectroscopy data do not allow us to resolve the photospheric nature of these targets, which would help strengthen the runaway hypothesis and allow us to dismiss the possibility that they are the result of strong ionizing radiation from the O-type population in NGC 2070.

Despite that fact that we cannot confirm the runaway hypothesis in these two cases, the detections obtained by MUSE are further supported by the corroborating data from NIRCam and MIRI. However, to gain a comprehensive understanding of the nature and kinematics of these objects, IR  spectroscopy is necessary. This study focused on the objects in fields A and B (Figs.~\ref{fig:area1} and \ref{fig:area2}), yet the electron density map of NGC\,2070 reveals several other compact regions of interest that may also be associated with forming stars or stellar interactions and/or mergers. The NGC 2070 field has an abundance of highly dense and compact sources, such as the object (or objects) located between fields A and B seen in Fig. \ref{fig:all}, which warrant a closer and more comprehensive examination through subsequent works involving deep IR imaging and spectroscopy.

The core of the Tarantula Nebula, NGC 2070, is a complex environment strongly shaped by the radiation from its rich population of massive stars \citep[e.g.,][]{2002AJ....124.1601W,2010ApJS..191..160P}. Within this maelstrom of ionized structures revealed in NGC 2070's ISM, the MUSE electron density map helps focus attention on points of interest where star formation or stellar interactions may be occurring. MUSE has identified these targets, and the IR integral field spectroscopic capabilities of NIRSpec on the JWST will be instrumental in mapping these regions and unveiling the detailed properties and motions of these sources in future studies.

\begin{acknowledgements}
        The author thanks the referee for useful comments and
        helpful suggestions that improved this manuscript. NC acknowledges funding from the Deutsche Forschungsgemeinschaft (DFG) - CA 2551/1-1,2 Our research used Astropy, a community-developed core Python
        package for Astronomy \citep{2013A&A...558A..33A}, and APLpy, an
        open-source plotting package for Python \citep{2012ascl.soft08017R}. The Early Release Observations and associated materials were developed, executed, and compiled by the ERO production team. The EROs were also made possible by the foundational efforts and support from the JWST instruments, STScI planning and scheduling, Data Management teams, and Office of Public Outreach. Credits: NASA, ESA, CSA, STScI, Webb ERO Production Team.
\end{acknowledgements}

\bibliographystyle{aa}
\bibliography{blobs_V1}

\end{document}